\documentclass{article}

\usepackage{mathrsfs}
\usepackage{amsfonts}
\usepackage{amsmath}
\usepackage{amssymb}
\usepackage{graphicx}

\title{Dosimetry, scattering theory, and Monte Carlo simulation}
\author{Gordon McCabe}

\def\eqalign#1{\,\vcenter{\openup.7ex\mathsurround=0pt
 \ialign{\strut\hfil$\displaystyle{##}$&$\displaystyle{{}##}$\hfil
 \crcr#1\crcr}}\,}

\begin{document}

\maketitle

\begin{abstract}
The purpose of this paper is to provide an introduction to the
physics of scattering theory, to define the dosimetric concept of
linear energy transfer in terms of scattering theory, and to provide
an introduction to the concepts underlying Monte Carlo simulations.

\end{abstract}

\section{Introduction}

Whilst the \textit{absorbed dose} deposited by energetic particles
(`radiation') in condensed matter is simply the energy per unit mass
delivered to the medium, the \textit{equivalent dose} is a function
of both the energy per unit mass, and the spatial distribution with
which that energy is deposited. The greater the spatial concentration
of deposited energy, the greater the equivalent dose. The average
rate at which a type of radiation locally deposits energy in a
medium, per unit distance of track length, is called the
\textit{linear energy transfer} (LET) of that radiation in that
medium. Radiation with higher linear energy transfer deposits a
greater equivalent dose in a medium.

The LET is closely related to the \textit{stopping power} of the
medium, which is the energy lost by an incoming particle, whether it
is deposited locally, or transformed into other radiation, such as
Bremsstrahlung photons, or the energy of secondary electrons,
(so-called `delta-rays'). Linear energy transfer and stopping power
generally coincide in the case of heavy charged particles, but the
linear energy transfer of beta-radiation (i.e, incoming electrons)
does not include the energy transformed into Bremsstrahlung photons
or delta-rays.

The first purpose of this paper is to define linear energy transfer
in terms of the concepts used in scattering theory, and, in
particular, the concepts of quantum scattering theory. An exposition
will then be provided of the principles underlying Monte Carlo
simulations.

\section{Scattering}

Scattering theory is the branch of physics which describes the
collision processes between particles. These are typically considered
to be processes in which an incoming particle, or beam of such
particles, interacts with a target particle, or collection of
particles. After an individual interaction, the energy and momentum
of both the incoming particle and the target particle can be altered.
A change of momentum includes both a change in the direction of
travel, and, in the case of particle with non-zero mass, a change in
the speed of travel. The target particle is often considered to be a
composite particle, and as such, both its energy of motion, and its
internal energy state, can change as a result of the collision.

Consider first the scattering of a particle beam in terms of
classical physics. In the case of a beam of particles with non-zero
mass, the beam has a flux $F$ defined as

$$
F = n_i v_i \;,
$$ where $n_i$ is the number of particles per unit volume in the
beam, and $v_i$ is the average velocity of the particles in the beam.
The flux has dimensions of $\text{Area}^{-1} \; \text{Time}^{-1}$. It
can be equivalently defined as the number of particles passing though
a unit cross-sectional area of the beam, per unit time. The flux
should not be confused with the \textit{fluence} $\Phi$ of a particle
beam, which is the number of particles passing through a unit
cross-sectional area of the beam, over the time duration considered.
Hence, the fluence is the definite integral of the flux over time,
$\Phi = \int_{t_0}^{t}F(\tau)\; d\tau$, and the flux is the time
derivative of the fluence, $F = d\Phi/dt$. The fluence has dimensions
of $\text{Area}^{-1}$.

A particle detector placed at an angle $\Omega=(\theta, \phi)$ with
respect to the direction of the incoming beam, will detect particles
at the rate (Bohm 1979, p312):

$$\eqalign{
N/\Delta t &= \sigma(\Omega) N_T n_i v_i \cr &= \sigma(\Omega) N_T F
\;.}$$ $N_T$ is the number of particles in the target, and the
constant of proportionality $\sigma(\Omega)$ is called the
differential cross-section\footnote{In the physics literature, the
differential cross-section is usually denoted as $d\sigma$, or
$d\sigma/d\Omega$. The expression for $d\sigma$ typically contains a
dependence upon both the direction and magnitude of the outgoing
momentum, whilst the expression for $d\sigma/d\Omega$ contains only a
dependence upon direction, and is obtained by integrating over all
the outgoing momenta in that direction. We shall refrain from such
mathematically questionable notation in this paper.} in the direction
$\Omega$. The differential cross-section can clearly be expressed as

$$
\sigma(\Omega) = \frac{1}{N_T} \frac{N/\Delta T}{n_i v_i}\;,
$$ or, in words,

$$
\sigma(\Omega) = \frac{1}{N_T} \frac{\text{Number of particles
scattered into }\Omega \; \text{per unit time}}{\text{Incident Flux}}
$$ The dimensions of the differential cross-section are
$\text{Area} \cdot \; \text{Solid Angle}^{-1}$.

Integrating the differential cross-section over all possible
directions gives the total scattering cross-section $\sigma$:

$$
\sigma = \int \sigma(\Omega) d\Omega \;,
$$ where $d\Omega = \sin \theta d\phi \; d\theta$.\footnote{In terms of differential
geometry, $d\Omega$ is the volume form on the 2-sphere, expressible
as $d\Omega = \sin \theta d\phi \wedge d\theta$, where $\wedge$ is
the antisymmetric tensor product.} The dimensions of the total
cross-section are simply $\text{Area}$. One can think of the
cross-section as effectively the area which the target presents to
the incoming beam.

In terms of the quantum mechanics of a scattering process, an
incoming particle is represented by a quantum state, or
`wave-function' $\Psi_i$, the interaction process is represented by a
scattering operator $S$, the final outgoing particle state is
represented by another wave-function $\Psi_f$, and the objective is
to calculate the transition probability $| \langle \Psi_f | S \Psi_i
\rangle |^2$ between an incoming state and an outgoing
state.\footnote{$\langle \cdot | \cdot\rangle$ here is the inner
product on the space of quantum states.} In many physically relevant
situations, the incoming state has a specific energy $E_i$ and
momentum $k_i$, and each possible outgoing state also has a specific
energy $E_f$ and momentum $k_f$. An outgoing state with a specific
momentum $k_f$, also has a specific direction $\Omega$ associated
with it.

The square-amplitude $|\Psi_i|^2$ of the incoming particle's
wave-function gives the position probability density for the
particle. In quantum mechanical terms the incoming flux $F_{\Psi_i}$
is then the product of the probability density per unit volume, with
the velocity of the particle. Equivalently, the quantum mechanical
flux is the probability density per unit area, per unit time. Hence,
in quantum mechanical terms even an individual particle has a flux
associated with it. Given the transition probability $| \langle
\Psi_f | S \Psi_i \rangle |^2$, there is also an associated
\textit{transition rate}, or transition probability per unit time. In
the simplest quantum mechanical terms, cross-sections can be defined
as follows (Bohm 1979, p314):

$$
\sigma = \frac{\text{Transition probability}}{\text{Incident
probability per unit area}} \;,
$$ or, equivalently,

$$ \sigma = \frac{\text{Transition probability,
per unit time}}{\text{Incident probability per unit area, per unit
time}} \;,
$$ which itself is equivalent to

$$
\sigma = \frac{\text{Transition rate}}{\text{Incident Flux}}\;.
$$

The quantum mechanical cross-sections used in practice provide an
economical way of bundling the transition probabilities between
classes of quantum states. For example, the differential
cross-section $\sigma(E,\Omega)$ is proportional to the probability
of a transition from \emph{any} incoming state $\Psi_i$ of energy $E$
to \emph{any} outgoing state $\Psi_f$ in which the momentum vector
$k_f$ points in the direction of $\Omega$. As before, integrating a
differential cross-section over all possible directions then gives a
total scattering cross-section $\sigma(E)$:

$$
\sigma(E) = \int \sigma(E,\Omega) d\Omega \;.
$$ This total cross-section is proportional to the probability of an incoming state of energy
$E$ interacting with the target particle.

Now, an \emph{elastic} collision is defined to be a collision in
which the outgoing particles are the same as the incoming particles,
and the total kinetic energy of the particles before the
collision is the same as the total kinetic energy of the
particles after the collision. Interactions in which this does not
occur are called \emph{inelastic} collisions. For example, an
inelastic collision occurs when an incoming particle interacts with
an atom, and some of the energy from the incoming particle excites or
ionises the atom. Alternatively, the incoming particle can be
absorbed by the target, raising the target particle into an excited
state, which may then subsequently release various decay products. An
example of this is the absorption of an incoming neutron by an atomic
nucleus. Hence, there are elastic scattering cross-sections,
inelastic scattering cross-sections, and absorption cross-sections,
all of which are functions of the energy of the incoming particle(s).

Given the inelastic differential cross-section $\sigma_\mu(\Omega)$
for an interaction in which the incoming projectile transfers $\Delta
E_\mu$ of energy to a target atom, and the projectile scatters with a
momentum $k_f$ in direction $\Omega$, the total inelastic
cross-section for all interactions in which the projectile transfers
$\Delta E_\mu$ of energy to a target atom is:

$$\eqalign{
\sigma_\mu &= \int \sigma_\mu(\Omega)d\Omega \cr &= \int_0^{2\pi}
d\phi \int_0^\pi \sigma_\mu(\theta,\phi)\sin \theta d\theta\;.}
$$

Given $N$ target atoms per unit volume in a condensed matter medium,
the product $N \sigma_\mu$ is the probability per unit path-length of
the incoming particle track, that the projectile will lose $\Delta
E_\mu$ of energy. Hence, the local energy transfer (linear energy
transfer) of the particle is given by the expression:

$$
-\frac{dE}{dx} = N \sum_\mu \Delta E_\mu \sigma_\mu.
$$ To be more precise, this is the quantum mechanical expectation
value\footnote{Given the probability $p_i$ for each possible value
$a_i$ of a quantity $A$, the expectation value $\langle A \rangle$ of
the quantity is defined to be the probability-weighted sum of those
values, $\langle A \rangle = \sum_i p_i a_i$.} for the energy loss
per unit length:

$$
- \left \langle \frac{dE}{dx} \right \rangle = N \sum_\mu \Delta
E_\mu \sigma_\mu \;.
$$

A special case of this equation is Bethe's formula for the local
energy loss, per unit track-length, of a swiftly moving charged
particle heavier than an electron. Bethe's formula applies, for
example, to energetic protons, alpha particles, and heavy ions. For
swift charged particles heavier than an electron, the inelastic
cross-sections are sharply peaked towards the forward direction. If
we take the forward direction to correspond to a polar angle of
$\theta = 0$, such particles scatter into a narrow cone-shaped region
circumscribed by a small polar angle $\Theta$, and all azimuthal
angles $\phi \in [0,2\pi)$. Bethe's formula then arises from assuming
that the cross-section

$$\sigma_\mu = \int_0^{2\pi} d\phi \int_0^\pi \sigma_\mu(\theta,\phi)\sin \theta
d\theta \;,
$$ can be approximated by

$$
\sigma_\mu = \int_0^{2\pi} d\phi \int_0^\Theta
\sigma_\mu(\theta,\phi)\sin \theta d\theta \;.
$$ Assuming the projectile has a non-relativistic speed, Bethe's formula
for linear energy transfer is:

$$
-\frac{dE}{dx} = 4\pi NZz^2 \frac{e^4}{m_e\nu^2}\ln\left(\frac{M\nu^2
\Theta}{I} \right)\;,
$$ where $M$ is the mass of projectile, $\nu$ is its speed, $ze$ is its charge, $m_e$ is
the electron mass, $Z$ is the proton number of the target atoms, $N$
is the number of target atoms per unit volume, and $I$ is a
parameter, called the `mean excitation potential', characterising
only the target atoms. Because the mass $M$ of the projectile only
appears in the natural logarithm, the charge of the incoming
projectile has a greater effect upon linear energy transfer than the
mass of the incoming projectile.

\section{Monte Carlo Simulations}

Monte Carlo simulations are computer simulations which use random
numbers. Given the inherently probabilistic nature of particle
interactions, Monte Carlo simulations can be used to simulate
particle tracks, and their local energy deposition in a chosen
medium. Monte Carlo simulations can thereby be used to calculate
absorbed dose and equivalent dose. The cross-sections obtained from
quantum mechanical and quantum field theoretical scattering theory
provide the probability distributions used in such Monte Carlo
simulations.

A computer can generate random numbers with any specified probability
distribution. In other words, a computer can generate numbers with
the relative frequencies specified by any probability distribution.
The ability of a computer to generate such random numbers is
dependent upon the ability of a computer to act as a `pseudo-random'
number generator. The computer implements an algorithm which
generates random numbers in the interval $(0,1)$ with an apparently
uniform distribution. The algorithm generates each number from its
predecessor according to a deterministic rule, hence the randomness
is purely one of appearance. The starting point of the pseudo-random
generator is called the `seed', and if one starts successive runs of
the program with the same seed, then exactly the same sequence of
numbers will be generated, in the same order, in each run. If one
wishes to avoid generating the same sequence of numbers on successive
runs, then one simply changes the seed of the generator. The
important point, however, is that the numbers are, to a specified
level of approximation, uniformly distributed in the interval
$(0,1)$.

Given the generation of random numbers with a uniform distribution in
$(0,1)$, random numbers can be generated in any other interval, with
any specified distribution. The method of doing so exploits the fact
that because the definite integral of any probability distribution
$p(x)$ must equal one,

$$\int^b_a p(x') dx' = 1 \;,$$ the indefinite integral $F(x)$, which uses the
upper limit of integration as the dependent variable,

$$
F(x) = \int_a^x p(x')dx'\;,
$$ must assume values in the interval between $0$ and $1$, beginning at the value of
$0$ when $x=a$, and increasing until it reaches the value of $1$ when
$x=b$.

Given a probability distribution $p(x)$ over a domain $(a,b) \subset
\mathbb{R}$, the selection of a random number $\mu$ between $0$ and
$1$, selects a value $\zeta \in (a,b)$ such that

$$F(\zeta) = \int_a^\zeta p(x')dx' = \mu \;.$$ When many values in
$(a,b)$ are selected by this means, they will approximate the
distribution specified by $p(x)$. The selection of random values
according to a specified probability distribution, in this manner, is
referred to as `sampling from a random distribution.'

In the Monte Carlo simulation of a particle track through a medium,
one typically generates random numbers for (i) the distance travelled
by the particle before its next collision; (ii) whether the next
collision is elastic or inelastic; (iii) the type of inelastic
collision in such an event; (iv) the energy transferred in the event
of an inelastic collision; and (v) the direction with which the
projectile scatters as a consequence of the collision. Let us look at
each of these in turn.

To calculate the distance travelled $l$ by a particle before its next
interaction, note first that where $\sigma_{\text{inelas}}$ denotes
the total inelastic scattering cross-section, and $N$ denotes the
number of targets per unit volume, the product
$N\sigma_{\text{inelas}}$ gives the probability per unit path-length
of the incoming particle track, that the projectile will undergo some
type of inelastic interaction. The reciprocal of this value therefore
gives the \emph{mean free path} $\lambda_{inelas}$ for inelastic
scattering. In other words, $\lambda_{inelas} = 1/N \sigma_{inelas}$
is the mean distance travelled by a particle between inelastic
collisions. Similarly, $\lambda_{elas} = N\sigma_{\text{elas}}$ gives
the probability per unit path length, that the projectile will
undergo some type of elastic interaction, and the reciprocal $1/N
\sigma_{elas}$ gives the mean free path for elastic scattering. The
total mean free path $\lambda_t$ is then

$$
\lambda_t = \frac{1}{\lambda_{inelas}^{-1}+ \lambda_{elas}^{-1}}\;.
$$

Now, the path-lengths between interactions will have the following
inverse-exponential distribution

$$p(x) = \frac{1}{\lambda_t} \exp \left(\frac{-x}{\lambda_t}\right)\;,$$ with
$\lambda_t$ as the mean of the distribution. Given a randomly
generated number $\mu_1 \in (0,1)$, the distance travelled $l$ by a
particle before its next interaction can be randomly generated
according to

$$
\mu_1 = \int_0^l \frac{1}{\lambda_t} \exp
\left(\frac{-x}{\lambda_t}\right) dx \;.
$$ The indefinite integral of the path-length distribution is

$$F(x) = \int \frac{1}{\lambda_t} \exp \left(\frac{-x'}{\lambda_t}\right)dx'  = -
\exp\left(\frac{-x}{\lambda_t}\right)\;,$$ hence

$$\eqalign{
\mu_1 &= \int_0^l \frac{1}{\lambda_t} \exp
\left(\frac{-x}{\lambda_t}\right)dx \cr &= F(l) - F(0) \cr &= -
\exp\left(\frac{-l}{\lambda_t}\right) + 1 \;.}
$$ Therefore,

$$1-\mu_1 = \exp\left(\frac{-l}{\lambda_t}\right)\;,$$ and

$$
\ln(1-\mu_1) = \frac{-l}{\lambda_t} \;,
$$ from which it follows that

$$
l = -\lambda_t \ln(1-\mu_1) \;.
$$ Given that $\mu_1$ is uniformly distributed in $(0,1)$, $1-\mu_1$ will also
be uniformly distributed in (0,1), hence, for the purposes of
generating random numbers, $\mu_1$ can be substituted into this
equation in place of $1-\mu_1$, to obtain the final expression:

$$
l = -\lambda_t \ln(\mu_1) \;.
$$ Given random numbers $\mu_1 \in (0,1)$, one uses this expression to generate
the random path lengths between collisions.

To decide whether the next collision is to be elastic or inelastic,
one can use the fact that the probability of an inelastic collision
$p_{inelas}$ is specified by

$$
p_{inelas} = \frac{\lambda^{-1}_{inelas}}{\lambda_{inelas}^{-1}+
\lambda_{elas}^{-1}} \;.
$$ A random number $\mu_2$ is generated, and if $\mu_2 \leq p_{inelas}$,
then the collision is chosen to be inelastic, otherwise it is chosen
to be elastic.

The types of possible inelastic collision then depend upon the type
of projectile under consideration. For example, in the case of an
incoming gamma-ray photon, there could be a photoelectric
interaction, a Compton-effect interaction, or a pair-production
interaction. The relative probabilities of the different interaction
types depend upon the energy of the incoming particle, and given
these relative probabilities, a randomly number $\mu_3 \in (0,1)$ can
determine which interaction is selected. Let $\kappa$ denote the
linear attenuation coefficient of energy $E$ gamma rays in the
medium, let $\tau$ denote the photoelectric attenuation coefficient,
and let $\varpi$ denote the Compton attenuation coefficient. If
$\mu_3$ is between 0 and $\tau/\kappa$, then photoelectric absorption
can be deemed to occur; if $\mu_3$ is between $\tau/\kappa$ and
$(\tau+\varpi)/\kappa$, then Compton scattering can be deemed to
occur; and if $\mu_3$ is between $(\tau+\varpi)/\kappa$ and 1, then
pair production can be deemed to occur (Mackie 1990, p552). The
energy lost as a result of the interaction is chosen, once more, by
the generation of random numbers, and by the energy loss
distributions specific to each type of interaction.

Given the relevant differential cross-section $\sigma(\theta,\phi)$
for the incoming energy $E$, random numbers can be generated which
specify the polar angle and azimuthal angle with which the projectile
scatters. Assuming the incoming projectile is travelling in the
direction $\theta = 0$, the probability $p(\theta)$ of scattering at
polar angle $\theta$ is

$$
p(\theta) = \frac{\sigma(\theta,\phi)\sin \theta}{\int_0^\pi
\sigma(\theta',\phi) \sin \theta' d\theta'} \;.
$$ The integral $\int_0^\pi
\sigma(\theta',\phi) \sin \theta' d\theta'$ of the differential
cross-section over all polar angles gives a quantity with dimensions
of $\text{Area} \cdot \; \text{Angle}^{-1}$, proportional to the
probability, per unit azimuthal angle, of scattering at any polar
angle. The expression $\sigma(\theta,\phi)\sin \theta$ gives a
quantity, with the same dimensions, proportional to the probability
of scattering, per unit azimuthal angle, at polar angle $\theta$. The
ratio of the latter by the former gives the probability $p(\theta)$
of scattering at polar angle $\theta$. Hence, given the selection of
a random number $\mu_4 \in (0,1)$, the angle $\theta$ such that

$$\eqalign{
\mu_4 &= \int_0^\theta p(\theta')d\theta' \cr &= \frac{\int_0^\theta
\sigma(\theta',\phi)\sin \theta'}{\int_0^\pi \sigma(\theta',\phi)
\sin \theta' d\theta'} }$$ is selected. The azimuthal angle $\phi$ is
determined by a uniform distribution over $(0,2\pi)$, and the
selection of another random number $\mu_5 \in (0,1)$.

\end{document}